# The rheology of ultra-high molecular weight poly(ethylene oxide) dispersed in a low molecular weight carrier 

Craig D. Mansfield,[a)] 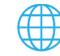 Tianran Chen (天然 陈), Mubashir Q. Ansari, 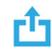 and Donald G. Baird[a)]

**AFFILIATIONS**

Department of Chemical Engineering, Virginia Tech, Blacksburg, Virginia 24061, USA

**Note:** This paper is part of the special topic, Celebration of Robert Byron Bird (1924-2020).
[a)]Authors to whom correspondence should be addressed: cdman@vt.edu and dbaird@vt.edu

## ABSTRACT

Gel spinning is the industrial method of choice for combining hydrophilic ultra-high molecular weight (UHMW) polymer resins with a hydrophobic support polymer to produce composite filaments for cytapheresis. Cytapheresis is a medical technique for removal of leukocytes from blood. Gel spinning is used to avoid high melt viscosity and thermal sensitivity of UHMW resins and the high melt temperature of the substrate resin but requires the recovery of toxic solvents. The UHMW resin is used because it forms a stable gel phase in the presence of water; a lower molecular weight resin (LMW) simply dissolves. UHMW and LMW resins were both poly(ethylene oxide) (PEO) and the substrate was polyarylsulfone (PAS). The literature indicated PEO undergoes non-oxidative thermal degradation above 200 °C and PAS is processed up to 350 °C. Dynamic oscillatory shear rheometry was used to study 0, 25, 40, 50, 60, and 75 wt. % UHMW PEO in LMW PEO to take advantage of the sensitivity of viscosity to changes in molecular weight and material configuration, indicating degradation. Samples were exposed to 220 °C, 230 °C, 240 °C, 250 °C, 275 °C, and 300 °C temperatures for 5 min to explore conditions that could result in sample degradation. The viscosity decreased less with increasing UHMW PEO content for samples exposed to the same temperature and the viscosity decreased more with increasing exposure temperature for samples with the same UHMW PEO content. Parameters were regressed from observed data to predict the change in molecular weight via empiricisms relating the viscosity to molecular weight, shear rate, temperature, and time.

## I. INTRODUCTION

The overall purpose of this investigation is to establish the conditions to be used for melt processing a powder blend of ultra-high molecular weight polyethylene oxide (UHMW PEO) in lower molecular weight polyethylene oxide (LMW PEO) as a part of a dual-extrusion process for combining UHMW PEO with a polyarylsulfone (PAS). UHMW resins are typically very difficult to melt process due to their extreme melt viscosities. A gel-based process, involving difficult and expensive solvent recapture, is the common solution when using UHMW resins in a flow process, and it is used to overcome wide differences in thermal sensitivity between components in composite materials. Similarly, dual-extrusion can be used to melt process materials with widely varying melting points and thermal sensitivities and form composites. This is typically achieved by taking advantage of physical phenomena such as subcooling of one or more melt streams.

The problems motivating the work performed are set up in the following description of the gel spinning process. The proposed change in the processing method is introduced with a description of the dual-extrusion process and then the materials involved and their selection are summarized. Finally, a discussion of rheology relevant to the experiments and data analysis ends this introduction with a brief summary of the information that should result.

The principle objective of this work is to determine the process conditions under which combining a LMW resin may allow an UHMW resin to be melt processed with minimal degradation at elevated temperatures. This can be split into questions that must be addressed by the work presented below: Does the addition of LMW resin to UHMW resin significantly lower the melt viscosity? How stable are the LMW and UHMW resins at elevated temperatures? What are the processing conditions required to minimize degradation?

## A. Gel spinning

Gel spinning is a polymer processing method wherein resins are solvated, extruded, and then treated to form the product.[1,2] Phase inversion gel spinning is a gel spinning method where the gel extrudate is immediately caused to precipitate the resin or resins using a bath. The solvents used to form the gel phase are generally toxic and must be removed completely from the filaments produced for medical applications.[3–5] Recapturing these toxic solvents typically introduces a significant overhead cost for this process.[6] A significant advantage is that a process designed around gel spinning typically removes the need to consider thermal incompatibilities when combining materials. Materials only need to be able to dissolve together in the same solvent. In this manner, the wide gap between the temperature above which the hydrophilic resin will degrade and the minimum temperature at which the rigid support material can be melt processed is overcome. This temperature mismatch is avoided completely with a solvent but at the cost of the solvent recapture overhead.

## B. Dual-extrusion

One method to limit the potential degradation of temperature sensitive resins is to plasticate the materials separately and combine them in a specialized mixing component. This is the basis for the dual-extrusion process.[7–9] In dual-extrusion, two extruders are used to separately plasticate the materials fed into the specialized mixer, in this case called a phase distributor. This provides better thermal control over the material streams and allows product designers to take advantage of physical phenomena, such as subcooling of one or more phases, to create novel interspersed materials. As dual-extrusion can be an entirely melt-based process, solvents can potentially be excluded. Careful consideration must be given when exploiting transient phenomena within the mixing component since the steady state conditions inside are potentially difficult to maintain or achieve. In the example of subcooling, a change in flow rate inside the mixing component can allow the materials sufficient time to freeze. For overcoming thermal incompatibility, this means that the properties of the material and how they change over time during elevated temperature exposure must be well characterized to determine the optimal processing conditions.

## C. Materials selection

When selecting materials for medical devices, emphasis is placed on low cost, wide use in industry, easy and economical processing, and a well-characterized safety profile. It is easier, and thus less costly, to seek approval for new medical devices when the materials involved have an established history of demonstrated safety and use in the context of the specific application.[10] The materials in the feed streams to the dual-extrusion phase distributor must be independently melt processable for it be a solvent-free process.

For the rigid support material, PASs are widely used in medical devices, with a demonstrated safety record. Udel, a PAS manufactured by Solvay, was selected for use due to its use in industry for gel spinning cytapheresis filaments. The principal disadvantage of using a PAS is the high temperature needed to melt process it. This is the reason a solvent-based process is generally preferred.[10,11]

There are several materials to choose from for the hydrophilic component. The candidate material currently used in industry for gel spun products is polyvinylpyrrolidone (PVP), commercially known as povidone. This material is well suited for a solvent-based process, but presents the potential for significant side reactions and cross-linking under elevated temperature conditions. In addition, the degradation byproducts are potentially hazardous. A better solution is the widely used polyethylene oxide (PEO). While not presently used in gel spun filaments for cytapheresis, PEOs have a well-established history of safe use and are also used in blood contacting applications. In addition, PEO resins generally degrade via chain scissioning reactions under non-oxidative thermolytic conditions, yielding only shorter chains and generally nontoxic byproducts that are easily removed by simply water extraction. Both PVP and PEO have molecular weight-dependent solubility in water, with ultra-high molecular weight resins forming a separate gel phase. It is this gel phase that is hypothesized to make cytapheresis filaments function. Unfortunately, ultra-high molecular weights also make melt processing very difficult since viscosity increases rapidly with molecular weight. To overcome the increased molecular weight required to form a stable gel phase for PEO, it is proposed that a lower molecular weight PEO can be used to disperse and carry the desired ultra-high molecular weight PEO through the process, to be removed with water in a later step.

## D. Rheology

In order to overcome the mismatch in temperature requirements, it is necessary to characterize the rheology of PEO at elevated temperatures and establish conditions for processing the material in dual-extrusion. A combination of simple rheological tests can yield the data necessary for analysis. To establish a baseline for the material viscosity, time–temperature superposition of dynamic oscillatory shear data at several temperatures can provide a reference curve at temperatures where transient behavior is not expected.[12] Stability of the melt phase can then be established at a generally accepted maximum processing temperature, $200\,°C$. Several thermogravimetric analysis and mass spectrometry studies have indicated that thermal degradation of PEO proceeds above $200\,°C$ via chain scissioning reactions at the C–O and C–C bonds along the chain.[13,14]

With an established stability, investigation of transient behavior at elevated temperatures can be performed with multi-frequency time sweeps. A first approximation involving a single polynomial term in the rate law is detailed in a description of the regression analysis found in the supplementary material. With the use of multiple temperatures and a first approximation assumption for the kinetics of changes in viscosity, the rate of change in viscosity and thus a presumed rate of change of average molecular weight can be predicted for any elevated temperature. The utility of such a first approximation will likely be limited outside the scope of this work, but will also better reflect the conditions of an industrial melt process. Direct translation of the diffusion limited degradation kinetics that describe a heated droplet to the approximately diffusion free bulk in pipe flow is at best non-trivial. Thermogravimetric studies suffer in this regard from the impact of a relatively large melt-atmosphere interface surface area relative to the sample droplet volume, which must be taken into account when selecting an appropriate rate equation.[14] In this work, the above three rheometric methods are combined into a single test, described later.

## II. EXPERIMENT
### A. Materials

There were two materials used for this work. The first material was a poly(ethylene oxide), designated as LMW PEO and sold by the Dow Chemical Company as the POLYOX WSR N10 product. It has a nominal $M_W$ of 100 000 g/mol. Per the sodium dodecyl sulfate (SDS), it contains 95 wt. % or greater of PEO, with the makeup consisting of silica, calcium carbonate, and less than 1 wt. % of the total as absorbed moisture. The second material was a poly(ethylene oxide), designated as UHMW PEO, sold by The Dow Chemical Company as the POLYOX WSR 303 product. It has a nominal $M_W$ of 7 000 000 g/mol. Per the SDS, it also contains 95 wt. % or greater of PEO, with the makeup consisting again of silica, calcium carbonate, and less than 1 wt. % of the total as absorbed moisture. Other work found in literature indicates that the distribution of molecular weights in the POLYOX and related series of PEO products available from The Dow Chemical company exhibit sharp peaks about the nominal average molecular weight.[15–17] It was assumed that this was the case for the provided materials.

Samples for experiments were prepared as follows. Batch mixing samples of the pure PEOs were collected in small aluminum dishes from their respective source containers and sealed within plastic bags for temporary storage to prevent moisture absorption. The required weights of the LMW PEO and UHMW PEO were added to a clean plastic bag from these batch mixing samples. The bag was chosen such that the volume was always significantly larger than the sample being prepared by a factor of approximately 30 or more. Dry nitrogen was used to inflate the bag and manual mixing was then performed for approximately 20 min. After mixing, the samples were then deposited in an aluminum dish and placed in a bag for temporary storage. This procedure was followed the morning of each day to minimize the influence of environmental factors in the lab. Samples were prepared for 0, 25, 40, 50, 60, and 75 wt. % of UHMW PEO in LWM PEO.

### B. Equipment

The rheometry was performed using a TA Instruments ARES G2 Rheometer. A 25 mm diameter stainless steel parallel plate fixture was used and experiments were performed under a dry nitrogen atmosphere. The rheometer was controlled via the TA Instruments Trios software.

### C. Preloaded rheometer test

A preloaded test procedure was constructed in the Trios software to automate the experimental work performed with the Ares G2 rheometer. This preloaded test procedure is explained in detail in the supplementary material. A brief summary of the preloaded test procedure and the components pertinent to this work follows.

Broadly, the full preloaded test procedure was broken into six parts and a total of 35 steps were involved. The data used to perform the regression analysis for this work were collected in parts two, three, and four. Further detailed discussion will be limited to these parts. Part 2 of the test consisted of steps 9 through 17 and involved generating the data required to construct a complex viscosity vs frequency master curve. Part 3 of the test consisted of step 18 and involved generating the data required to assess the stability of PEO at 200 °C before exposure to an elevated temperature. Part 4 of the test consisted of steps 19 through 23 and involved generating the data required to investigate how elevated temperatures affect PEO samples.

Steps 9 through 33 were performed either in isothermal frequency sweeps, isothermal multi-frequency time sweeps, or multi-frequency temperature ramps. While some parameters necessarily changed between steps of the same type, the common parameters can be summarized for the sake of brevity. The isothermal frequency sweep steps each began with a soak time of 30 s. The strain amplitude used was 0.5% and the frequency was logarithmically swept from 1 rad/s to 600 rad/s with 20 points/decade. The same multi-frequency strain signal was specified for all of the isothermal multi-frequency time sweeps and multi-frequency temperature ramps. The angular frequencies were 10, 20, 30, and 40 rad/s with strain amplitudes 0.5, 0.1, 0.1, and 0.1%, respectively. The sampling rate was set to 100 points/s for all multi-frequency steps and the soak times were set to 0 s. Steps 20 and 23 were the only isothermal multi-frequency time sweep steps with exit conditions specified. These steps were set to terminate once the temperature had equilibrated, with 20 consecutive points within a 0.125% tolerance. In practice, the exit conditions were met within, at most, several seconds.

In Table I is shown a reduced layout of the preloaded rheometer test with relevant parameters given for the indicated step types. Only the steps in the test that produced data used in the regression analysis are shown. The complete version of this table can be found in the supplementary material as Table I.

**TABLE I.** Reduced layout of preloaded rheometer test. IFS, isothermal frequency sweep; MFTS, multi-frequency time sweep.

| Part | Step | Type | Start temp (°C) | End temp (°C) | Temp ramp rate (°C/min) | Sweep time (s) | Frequencies (rad/s) |
|---|---|---|---|---|---|---|---|
| 2 | 9 | IFS | 100 | | | | 1–600[a] |
| | 11 | IFS | 125 | | | | 1–600[a] |
| | 13 | IFS | 150 | | | | 1–600[a] |
| | 15 | IFS | 175 | | | | 1–600[a] |
| | 17 | IFS | 200 | | | | 1–600[a] |
| 3 | 18 | MFTS | 200 | | | 1200 | 10, 20, 30, 40 |
| 4 | 21 | MFTS | Peak | | | 300 | 10, 20, 30, 40 |

[a]20 points/decade.

## D. Procedure

The rheometer was prepared with the fixture and sample ring in place. The normal setup was performed for a parallel plate test with a preheat to 100 °C. The preloaded test procedure described above was then adjusted for the peak temperature for the particular test. When the rheometer was thermally stabilized and ready to continue, the aluminum dish containing the prepared powder blend of UHMW PEO and LMW PEO was removed from the plastic bag. From this blend, a $1.00 \pm 0.01$ g sample was collected in a scoopula and then loaded into the rheometer with the fixture gap set to 6 mm. The rheometer was closed and the previously described preloaded test procedure in the software was started. At designated pause points, the rheometer was opened to lower the sample ring and to trim the sample, respectively. Excess sample material was quickly removed from the outside of the fixture. Upon test completion, the top half of the fixture was detached from the rheometer and the rheometer was opened to retrieve the sample. The sample was carefully removed as an intact disk, if possible, and placed in a marked aluminum dish. The warm fixture was cleaned and the rheometer was reset.

## III. DATA AND ANALYSIS

The principal goal of the data analysis was to extract and predict changes in the average molecular weight using viscosity as a proxy. Several empirical relationships were necessary to achieve this. A more complete description of the following empirical construction is given in the supplementary material. The generalized van der Waals equation of state with modification for the ambient pressure was selected to provide a means of relating the temperature and density of the pure PEO materials.[18] A modification to the Carreau–Yasuda viscosity function was adopted according to similar modifications made for the Cross and Carreau relationships in the literature,[19] with the lambda parameter replaced by the zero shear viscosity normalized by a new characteristic shear stress parameter. The modified viscosity function is given as

$$\frac{\eta}{\eta_0} = \left(1 + \left(\frac{\eta_0}{\tau_*}\omega\right)^{2a}\right)^{\frac{n-1}{2a}}, \quad (1)$$

where $\eta$ is the viscosity, $\omega$ is the angular frequency, $\eta_0$ is the zero shear viscosity, $\eta_\infty$ is the infinite shear viscosity, $\tau_*$ is the characteristic shear stress parameter, $n$ is the power law index, and $a$ is a dimensionless parameter. The infinite shear viscosity (not shown) was assumed to be zero for ease of computation. The Cox–Merz rule was assumed to hold for this work, so the angular frequency and shear rate were assumed to be interchangeable. Since the experiments were carried out well above the literature glass transition temperature range for PEO,[15] the Arrhenius form for horizontal shift factors was used to perform time–temperature superposition. The corresponding vertical shift factor was defined by the conventional temperature and density dependence.[12] The zero shear viscosity was coupled to the molecular weight using the power law empiricism of Teramoto and Fujita.[20] The shift factors and relation to the viscosity are given as follows:

$$\eta(M_R, T_R, \omega) = \eta(M, T, a(M, T)\omega)\frac{b(M, T)}{a(M, T)}, \quad (2)$$

$$b(M, T) = \frac{\rho(M, T_R)T_R}{\rho(M, T)T}, \quad (3)$$

$$a(M, T) = \left(\frac{M}{M_R}\right)^\alpha \exp\left(-\frac{E_{a,\nu}}{R}\left(\frac{1}{T} - \frac{1}{T_R}\right)\right), \quad (4)$$

where $a$ is the horizontal shift factor, $b$ is the vertical shift factor, $\rho$ is the density, $E_{a,\nu}$ is the Arrhenius activation energy, $R$ is the ideal gas constant, $M$ is the molecular weight, $T$ is the temperature, and the subscript $R$ denotes the reference conditions of the master curve. A first approximation for the time dependence for the molecular weight was assumed, given the presence of only chain scissioning reactions from literature studies.[13,14,21] The rate of change of the average molecular weight was expressed as proportional to a power of the average molecular weight. The rate coefficient was expressed in an Arrhenius form and the trivial integrated forms for the "extent of reaction" were expressed as functions of time and via the rate coefficient, temperature. The composite nature of the samples was used to separate and regress the pure UHMW PEO viscosity from the sample viscosity by assuming the conventional rule of mixture by volume for material properties and regressing the requisite exponent. For the analysis performed in this work, the reference temperature was set to 100 °C and the reference molecular weight was set to that of the LMW PEO, 100 000 g/mol. The choices of reference temperature and molecular weight are, in principle, arbitrary and were selected to coincide with the first temperature at which the observation of samples began and the MW of the pure LMW PEO. As a practical matter, these reference conditions were chosen to coincide with the initial conditions of the experimental work. This reduced the computational workload as the shift factors for data generated at the reference conditions are equal to unity.

The frequency sweeps from part 2 of the rheometry test were fit to form a dimensionless master curve using the GRG Nonlinear Solver in the Microsoft Excel software. It was assumed that the sample remained a two-phase composite during part 2 of the test. The viscosity of the LMW PEO and UHMW PEO phases were predicted and then combined to predict the composite sample melt viscosity using the previously described empiricism. The error between the prediction and the observation was then minimized using the GRG Nonlinear Solver to regress parameters. This characterized the dependence of melt viscosity on composition, molecular weight, temperature, and angular frequency, leaving the parameters describing time dependence to be regressed from other data.

The stability of PEO was assessed from data collected in part 3 of the test by calculating the relative deviation of the sample viscosity from the viscosity at the beginning of part 3 of the test. As this was an otherwise qualitative determination, no further numerical analysis was required and additional discussion is given with the results below.

Using the parameters regressed above from part 2 of the test, it was possible to regress the remaining parameters describing the assumed time dependence of the average molecular weight from data collected in part 4 of the test, using the melt viscosity as a proxy. Similar to above, the change in melt viscosity relative to the start of the isothermal multi-frequency time sweep in test step 21 was predicted using the constructed empiricism and the error relative to the observed viscosity was then minimized using the same GRG Nonlinear Solver in the Microsoft Excel software. This yielded the rate coefficient and extent of reaction for each test. The best convergence was achieved with the assumption of a first order rate law. The rate coefficient was then averaged over the compositions tested at each "peak temperature" under the assumption that the activation energy was not dependent on the average molecular weight. The averaged rate coefficient was then regressed with respect to the inverse temperature relative to the reference temperature to produce the pre-exponential coefficient and the activation energy, completing the set of regressed parameters.

Analysis of data from later in the tests performed was not necessary for this work and was not carried out. Construction of plots for process parameter selection was performed by specifying the temperature and angular frequency, then plotting the viscosity over time for a range of compositions. There are many ways to plot the 5D data related by the constructed regression. This representation was chosen to allow relevant process and product constraints to be easily included, which is commented on further in the Results and Discussion section below.

## IV. RESULTS AND DISCUSSION

### A. Base rheology of PEO

The base (pre-exposure) rheological behavior of PEO melts under dynamic mode small amplitude oscillatory shear is exhibited first. In Fig. 1 is shown a plot of complex viscosity, $|\eta^*|$ (Pa·s), vs angular frequency, $\omega$ (rad/s), for various compositions of PEO at 200 °C before exposure to elevated temperature as individual master curves with 95% confidence level error bars for each respective composition. Of particular note is the significant difference in behavior between the pure LMW PEO and the samples containing the UHMW PEO. The LMW PEO curve exhibits a clear tendency toward a finite zero shear viscosity in the low shear region whereas the presence of UHMW PEO results in a curve more characteristic of a power law relationship between $|\eta^*|$ and $\omega$. The curve corresponding to the lowest concentration of UHMW PEO tested, 25 wt. %, still exhibits some of the curvature present from the LMW PEO present in the low shear region. However, this influence is virtually imperceptible for the remaining curves. This indicates that the UHMW PEO content strongly affects both the magnitude and shear dependence of the composite viscosity. This is further demonstrated by the increase in slope with concentration observed for the 50, 60, and 75 wt. % compositions. Of particular note is the apparent crossing of the 50 and 60 wt. % curves at low frequency. While no statistically significant difference in viscosity between 50 and 60 wt. % compositions was observed, the mean viscosity for the 50 wt. % material remained above that of the 60 wt. % material after the apparent crossing. A similar condition exists at high frequency between viscosity curves for the 50 and 75 wt. % compositions. This reversal in magnitude ordering, relative to composition, with increasing frequency is a simple expected geometric consequence for materials that exhibit such power law behavior. Observation of these apparent intersections between viscosity curves without *a priori* knowledge of their locations occurred by simple coincidence. They are to be expected over a sufficiently large frequency range.

In Fig. 2 is shown the master curve plot of reduced complex viscosity, $|\eta^*|/\eta_0$, vs reduced angular frequency, $\lambda\omega$. The base rheological behavior presented in this plot was produced by regressing the pure UHMW PEO viscosity from the composite data, shifting all with respect to molecular weight to collapse to a single curve, and then normalizing by the appropriate parameters to produce dimensionless data. The LMW PEO and UHMW PEO data are shown with 95% confidence level error bars, and the regression curve is included to show the corresponding trend. The regressed reference point zero shear viscosity, $\eta_{0,R}$, was found to be 50.0 KPa·s. The regressed characteristic stress parameter, $\tau_*$, was 4.68 MPa. The regressed index parameter, $n$, was 0 and the regressed exponent parameter, $a$, was 0.0902. The regressed activation energy, $E_{a,v}$, for the temperature dependence of the vertical shift was −35.3 KJ/mol·K and the exponent for the molecular weight dependence, α, was 3.4.

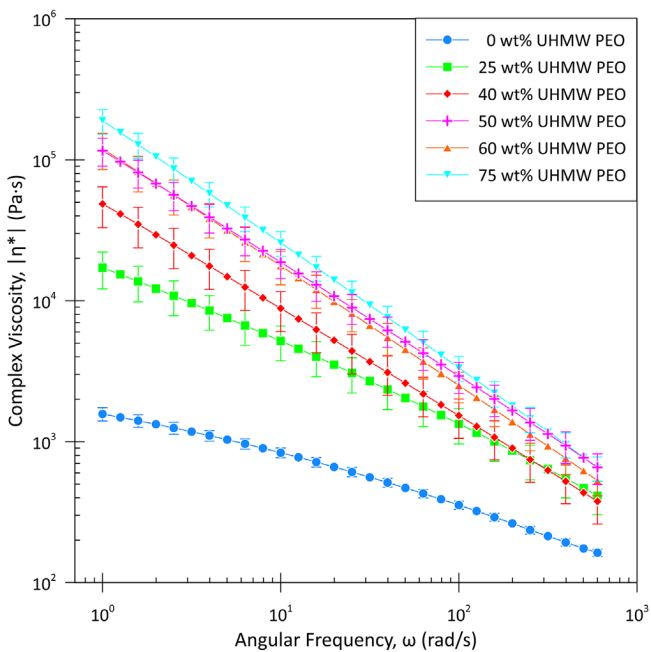

**FIG. 1.** Pre-exposure viscosity of PEO at 200 °C, complex viscosity, $|\eta^*|$ (Pa·s), vs angular frequency, $\omega$ (rad/s).

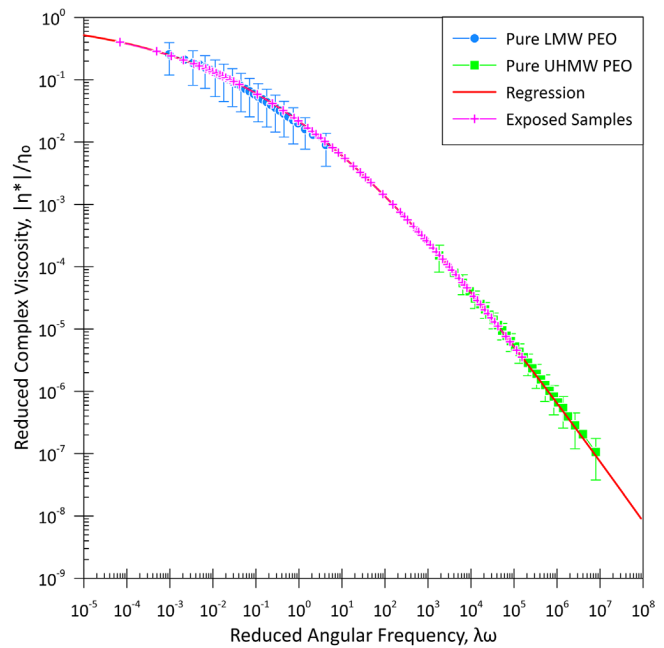

**FIG. 2.** Reduced variables master curve regression for PEO from pre-exposure and exposure data, reduced complex viscosity, $|\eta^*|/\eta_0$, vs reduced angular frequency, $\lambda\omega$.

While this figure is necessarily referenced to 100 °C and 100 000 g/mol by the reference conditions used to perform the regression analysis, it is not itself referenced to a particular set of conditions since specifying any zero shear viscosity and any relaxation time will yield a new plot specific to those conditions for PEO. It is a material dependent curve representative of the behavior of PEO for any temperature and any molecular weight. The independence of this master curve from temperature and molecular weight and the material dependence of the fit parameters defining its shape provide a critical means of relating changes in these conditions.

In addition, data produced during the "peak temperature" exposure in step 21 of the previously described test is shown in Fig. 2. The data were shifted with respect to the measured temperature and the average molecular weight predicted from the above regressed parameters and the measured complex viscosities. The data collected during elevated temperature exposure shifted readily to match the master curve for all samples tested. Regardless of the initial position of a sample material on the master curve, all measured samples traversed the curve from right to left with respect to increasing time, corresponding to a change in the average molecular weight as predicted by the regression performed above. In this way, the master curve presented also includes a means of predicting the change in complex viscosity with respect to time without the need to explicitly measure or calculate the average molecular weight of a sample of PEO. Practical utility of a single, material dependent master curve connecting changes in viscosity to changes in shear rate, temperature, time, and molecular weight will be demonstrated later in this work. However, it is not possible to calculate a predicted shift with respect to time along the master curve without some knowledge of the kinetics governing the changes in the material while exposed to elevated temperatures.

### B. Degradation of PEO

Before exploring the kinetics for changes in the material, change ought to be observed. In Fig. 3 is shown a plot of complex viscosity, $|\eta^*|$ (Pa·s), vs time, t (s), that demonstrates the change in the measured complex viscosity and regressed zero shear viscosity of 75 wt. % UHMW PEO in LMW PEO over 5 min when exposed to 300 °C. The measured complex viscosity data come directly from the isothermal multi-frequency time sweep performed in part 21 of the previously described test and the corresponding zero shear viscosity from the regression analysis. These conditions were selected to demonstrate the clearest example of the change in viscosity that occurs during an elevated temperature exposure, and the zero shear viscosity was included to demonstrate the greatest extent of change predicted to be possible for the sample. The zero shear viscosity provides an upper bound for the sample viscosity. The large decrease over time indicates that a corresponding substantial change in molecular weight is predicted to have occurred, independent of angular frequency. The implied decrease in molecular weight with decrease in viscosity over time at constant temperature similarly implies the presence of the above-described degradation found in literature.

A direct measurement of the sample molecular weight distribution after exposure to elevated temperature could plausibly confirm the form and extent of degradation. However, the extreme range of molecular weights to be measured presents a unique challenge with standard equipment and tests, which was not possible to overcome within the scope of this work. The standard gel permeation chromatography (GPC) method quickly becomes impracticable from either the extreme column length or the low flow rate and analyte concentration required to perform a successful measurement. As such, the implied changes in molecular weight remain implied.

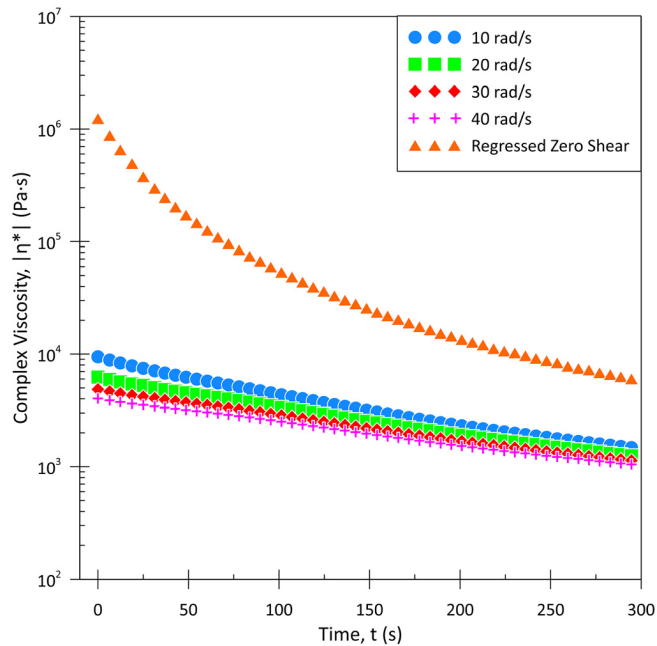

**FIG. 3.** Viscosity of 75 wt. % UHMW PEO in LMW PEO during 5 min exposure to 300 °C, measured complex viscosity and regressed zero shear viscosity, complex viscosity, $|\eta^*|$ (Pa·s), vs time, t (s).

### C. Kinetics of PEO at elevated temperature

In Fig. 4 is shown the Arrhenius plot for the regressed rate of change coefficients as rate of change coefficient, k (mol/g·s), vs inverse temperature, 1/T (K$^{-1}$), with 95% confidence level error bars. A line representing the calculated rate of change coefficient corresponding to the parameters regressed from the data is also shown. From simple linear regression, the activation energy, $E_{a,x}$, was (120 ± 11) KJ/mol·K and the natural log of the pre-exponential coefficient, $\ln(k_0)$, was (−19.0 ± 1.1). The coefficient of determination for the regression, $r^2$, was 0.983, which indicated a good fit of the data. The data for this plot were regressed from the data collected from the isothermal multi-frequency time sweeps performed at the "peak temperatures" during step 21 of the previously described test. From the zero shear viscosity and the relaxation time regressed at each point in time, the relative changes in predicted average molecular weight and extent of reaction were calculated using the previously regressed parameters describing the base rheology of PEO and then regressed to yield the rate of change coefficient. With the assumption that the activation energy of the change in viscosity, $E_{a,x}$, is independent of molecular weight, the rate of change coefficient was then averaged over sample composition.

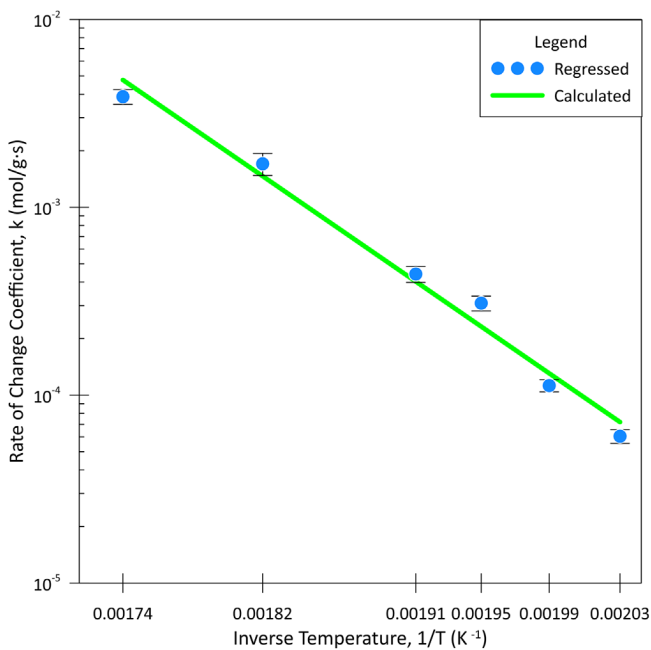

**FIG. 4.** Arrhenius plot of the first order rate of change coefficient, averaged over composition at constant temperature, rate of change coefficient, $k$ (mol/g·s), vs inverse temperature, $1/T$ (K$^{-1}$).

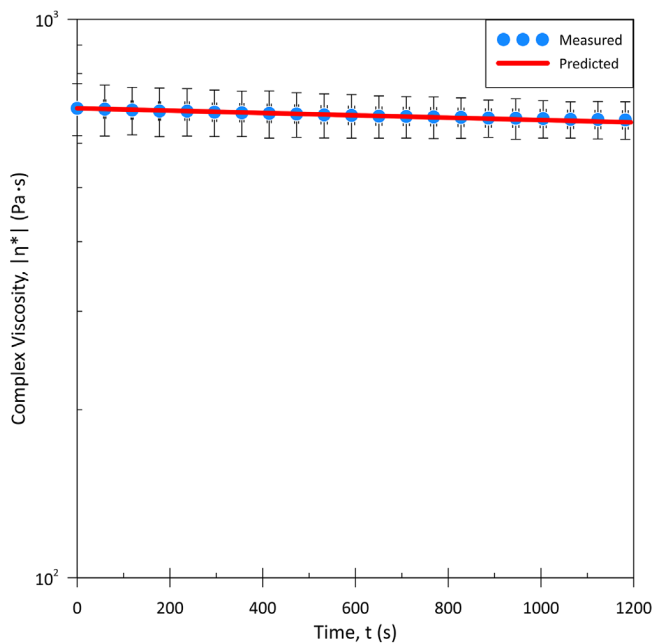

**FIG. 5.** Pre-exposure stability of LMW PEO at 10 rad/s at 200 °C over 20 min, comparison of measured viscosity with prediction from regression, complex viscosity, $|\eta^*|$ (Pa·s), vs time, $t$ (s).

These regressed rate of change coefficients correspond to the initial sample composition and "peak temperature" defining the test conditions. It was found during the regression of the rate of change coefficients that the additional parameter describing deviation from a first order dependence on the average molecular weight, $a_r$, invariably tended toward zero. This reduction to zero was independent of the sign of the initial value of the parameter and it indicated that the best fit for the observed behavior is a first order rate of change. With these parameters regressed, the set of parameters required to predict the viscosity, as previously described, is complete and some additional commentary on the stability of PEO at 200 °C can be made.

### D. Stability of PEO

In Fig. 5 is shown a plot of complex viscosity, $|\eta^*|$ (Pa·s), vs time, $t$ (s), describing the observed pre-exposure stability of LMW PEO at 10 rad/s at 200 °C over 20 min and a line showing the prediction from regression. The data, collected during part 3 of the previously described test, are shown with 95% confidence level error bars. At this level of confidence, no statistically significant change in viscosity was observed for the LMW PEO samples tested. However, the mean measured viscosity did decline approximately 5% over the course of 20 min with respect to the initial time, indicating that stability beyond 20 min should not be assumed without further study. This composition was selected as it exhibited the greatest relative change in mean measured viscosity. All other compositions tested similarly exhibited no statistically significant change in viscosity. This indicates that the materials tested are remarkably stable at this temperature, but could eventually exhibit a significant change in viscosity beyond 20 min, indicating degradation. The predicted initial average molecular weight was 95 158 g/mol and it was calculated by matching the value of the viscosity at the first data point. The predicted viscosity shown is in good agreement with the mean value trend, deviating by no more than approximately 0.9%, and is shown to fall well within the error bounds of the measured data. From this additional information, it can be concluded that no change in sample viscosity was observed over the 20 min held at 200 °C and that the viscosity of PEO is predicted to deviate by no more than approximately 1.5% after 5 min and 5.6% after 20 min at 10 rad/s angular frequency. The lowest angular frequency used in the multi-frequency steps of the previously described test was 10 rad/s. Since lower shear rates generally respond more strongly to changes in the average molecular weight, it can further be stated that these predicted maximum deviations are also generally expected to be true for any angular frequency greater than 10 rad/s.

Predictions made from the parameters regressed from separate parts of the test, parts 2 and 4, were in agreement with an intermediate and separate part of the test, part 3, with a maximum deviation of 0.9% from the mean observed viscosity and no deviation from agreement at the 95% confidence level. This indicates that the previously described empirical construction may be used to predict the complex viscosity of pure and composite PEO materials within the range of compositions, molecular weights, temperatures, times, and angular frequencies included in this work.

### E. Process parameter selection for dual-extrusion

The predictions made using the previously described empirical construction using the regressed parameters represent a five-dimensional relationship with complex viscosity as a function of time, temperature, angular frequency, and initial composition or molecular weight. The relationship between complex viscosity and time is what will ultimately drive the selection of process parameters, and it presents an obvious choice of axes for a useful plot. This choice of axes leaves three degrees of freedom remaining, one of which is still reasonable to leave unspecified to produce a 3D dataset. There are advantages for including contour curves for temperature, angular frequency, and initial composition in respective cuts along the parameter space. Given this complexity and the similarity of the resulting use of the generated plots, further discussion will be limited to the case of initial composition contours.

In Fig. 6 is shown an example plot of complex viscosity, $|\eta^*|$ (Pa·s), vs time, t (s), which can be used for process parameter selection. To generate the plot, the remaining two variables—temperature and angular frequency—were specified as 275 °C and 100 rad/s, respectively. The boundary of the operable region of the parameter space may be defined by identifying the appropriate boundary conditions defined by the constraints of the process and the desired properties of the product. Lines representing the maximum allowed process viscosity, the minimum allowed product viscosity, and a nominal residence time in the process define a patch of allowed behavior. The lines that are able to remain within the defined viscosity boundaries for the entire residence time represent the allowed range of compositions for the specified temperature and angular frequency. Due to the need to specify these two remaining degrees of freedom, there is no guarantee that optimum, or any, allowed process conditions will be identified without additional information from the design of the process and the nominal constraints for the product.

## V. CONCLUSIONS

The principle objective for this work is to determine the process conditions under which combining a LMW resin may allow an UHMW resin to be melt processed with minimal degradation at elevated temperatures. This can be split into questions that must be addressed by the work presented above: Does the addition of LMW resin to UHMW resin significantly lower the melt viscosity? How stable are the LMW and UHMW resins at elevated temperatures? What are the processing conditions required to minimize degradation?

The first question can be answered by direct inspection of Figs. 1 and 2. A significant difference in both the magnitude and shear dependence of viscosity was observed between pure LMW PEO samples and those with UHMW PEO, indicating that a combination of composition and temperature can be used to adjust and lower the melt viscosity relative to pure UHMW PEO.

The second question can be answered from both observation and regression results. No change in viscosity was observed over 20 min for samples held at 200 °C (Fig. 5), and change was observed for all samples over 5 min for temperatures of 220 °C and above, with the greatest change for 75 wt. % UHMW PEO in LMW PEO at 300 °C (Fig. 3). The good agreement between observation and regressed prediction (Fig. 5) shows that the regressed description of viscosity time dependence is capable of extrapolation to observed stable conditions. The time dependence of viscosity at elevated temperatures is well described, even by a simple first order rate equation, via the dependence of zero shear viscosity on the calculated average molecular weight. It must be stressed again that this is a calculated value and not a measured average molecular weight.

The third question can be answered directly from the example usage of the regression results shown in Fig. 6. The viscosity is predicted and observed to depend on the shear rate, composition, temperature, and time. With four degrees of freedom available, the minimization of degradation is largely dependent upon constraints determined by the process and product. However, some logical deductions can be made. The relative decrease in viscosity over time is predicted to be less with increasing concentration of UHMW PEO in LMW PEO. With the example shown in Fig. 6, this is simply the corresponding line that remains between the viscosity constraints for the full residence time. Since the eventual goal of this work is to overcome a temperature mismatch, it can be concluded that the separate plastication of PEO from the other component, PAS, found in dual-extrusion would be better than combining all of the materials in a single extruder. The shorter residence time required at elevated temperatures would allow reduction of the extent of degradation by process design alone.

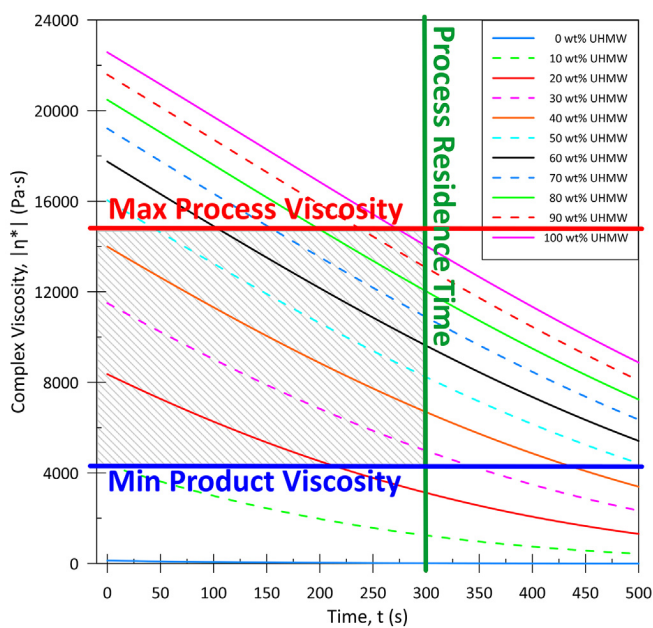

FIG. 6. Process parameter selection from predicted complex viscosity with example process and product constraints, exposure to 275 °C with 100 rad/s angular frequency, complex viscosity, $|\eta^*|$ (Pa·s), vs time, t (s).

## SUPPLEMENTARY MATERIAL

See the supplementary material for more information about the regression analysis and the preloaded rheometer test.

## ACKNOWLEDGMENTS

We acknowledge the financial support from Innovative Biotherapies provided by NIH NCATS through the SBIR Research Grant No. R43 TR001324–01.

## AUTHOR DECLARATIONS
### Conflict of Interest

The authors have no conflicts of interest to disclose.

## DATA AVAILABILITY

The data that support the findings of this study are available from the corresponding authors upon reasonable request.